\documentclass[3p]{elsarticle}
\usepackage[T1]{fontenc}
\usepackage[utf8]{inputenc}
\usepackage{lmodern}
\usepackage{epsfig} 
\usepackage{graphicx}
\usepackage{amsmath}
\usepackage{amssymb}
\usepackage{ulem}
\usepackage{color}
\usepackage{cleveref}
\usepackage{mathrsfs}
\usepackage{slashed}
\usepackage{subfig}
\usepackage{cuted}
\usepackage{mathtools}
\usepackage{lipsum}
\usepackage[compat=1.0.0]{tikz-feynman}

\tikzfeynmanset{
	fermion2/.style={
		/tikz/postaction={
			/tikz/decorate=false,
		},
	},
}
\tikzfeynmanset{
	fermion3/.style={
		/tikz/decoration={
			markings,
			mark=at position 0.5 with {
				\node[
				dot,
				fill,
				draw=none,
				] { };
			},
		},
		/tikz/postaction={
			/tikz/decorate=true,
		},
	},
}

\DeclareMathOperator*{\SumInt}{%
	\mathchoice%
	{\ooalign{\raisebox{.15\height}{\scalebox{0.9}{$\textstyle\sum$}}\cr\hidewidth$\displaystyle\int$\hidewidth\cr}}
	{\ooalign{\raisebox{.14\height}{\scalebox{.7}{$\textstyle\sum$}}\cr\hidewidth$\textstyle\int$\hidewidth\cr}}
	{\ooalign{\raisebox{.2\height}{\scalebox{.6}{$\scriptstyle\sum$}}\cr$\scriptstyle\int$\cr}}
	{\ooalign{\raisebox{.2\height}{\scalebox{.6}{$\scriptstyle\sum$}}\cr$\scriptstyle\int$\cr}}
}

\newcommand{\avg}[1]{\left\langle #1 \right\rangle}

\newcommand{\Tr}{\textrm{Tr}}

\definecolor{dgreen}{rgb}{0.1,0.5,0.1}
\definecolor{lblue}{rgb}{0.2,0.35,1}
\definecolor{webred}{rgb}{0.75,0,0}

\title{Inhomogenuous instabilities at large chemical potential \\ in a rainbow-ladder QCD model}

\address[1]{Institut für Theoretische Physik, \\ Justus-Liebig-Universität Gie\ss en,
	35392 Gie\ss en, Germany.\\ \,}
\address[2]{Technische Universität Darmstadt, Fachbereich Physik, Institut für Kernphysik, \\ Theoriezentrum, Schlossgartenstr.~2, 64289
	Darmstadt, Germany.\\ \,}

\address[3]{Helmholtz Forschungsakademie Hessen f\"{u}r FAIR (HFHF), \\
	GSI Helmholtzzentrum f\"{u}r Schwerionenforschung,\\
	Campus Darmstadt,
	64289 Darmstadt,
	Germany. \\ \,}
\address[4]{Helmholtz Forschungsakademie Hessen f\"{u}r FAIR (HFHF), \\
	GSI Helmholtzzentrum f\"{u}r Schwerionenforschung,\\
	Campus Gie\ss{}en,
	35392 Gie\ss{}en,
	Germany.}

\author[1,2]{Theo F. Motta}
\author[1,4]{Julian Bernhardt}
\author[2,3]{Michael Buballa}
\author[1,4]{Christian S. Fischer}

\begin{document}
	
	\begin{abstract}
        In this work we continue our efforts to study the existence of a phase with an inhomogeneous, i.e., spatially varying, chiral condensate
		in QCD. To this end we employ a previously established method of stability analysis of the two-particle
		irreducible effective action in a truncation that corresponds to a rainbow-ladder approximation of the quark-gluon 
		interaction of QCD. 
        If the analysis is restricted to homogeneous phases, the phase diagram features a first-order chiral transition in the lower-temperature regime.
        Performing the stability analysis along the lower-chemical-potential border of the corresponding spinodal region, we find that below a certain temperature the homogeneous chirally symmetric solution is unstable against inhomogeneous condensation. 
        We argue that this instability may persist to chemical potentials above the homogeneous first-order phase boundary, in which case it signals the existence of an inhomogeneous ground state.   
        Our methodology is also applicable for more sophisticated truncations of the QCD effective action.

	\end{abstract}
	
	\maketitle
	
	\section{Introduction}\label{sec:intro}
	
	It is still an open question if phases characterized by a non-zero spatially varying chiral condensate do exist in the QCD phase diagram.
	These inhomogeneous or crystalline phases may appear at large chemical potential. In this region of the QCD phase 
	diagram	pairing of opposite-handed quarks with opposite momentum into a scalar quark-condensate becomes energetically 
	disfavoured with respect to a chirally symmetric phase. On the other hand, it has been suggested that pairings with 
	non-vanishing total momenta might still be possible, resulting in an inhomogeneous chiral condensate \cite{Deryagin:1992rw,Park_2000,Rapp_2001,Nickel:2009wj,Kojo_2010}, see also Ref.~\cite{Buballa:2014tba}
	for a review. 
	
	In recent years, much has been learned from lower-dimensional models. The existence of a crystalline phase has been 
	unequivocally proven for the $1+1$ dimensional Gross-Neveu (GN) model at low temperatures and intermediary to high 
	chemical potential in the limit of infinite fermion flavours $N$ \cite{Thies:2006ti,Koenigstein:2021llr}
    as well as for finite $N$ at zero temperature \cite{Ciccone_2022}. 
 For higher dimensions a systematic survey of various four-fermion and Yukawa models has revealed the following pattern:
 For $d+1$ space-time dimensions the inhomogeneous phase is present for 
	$d<2$ (see Refs.~\cite{Koenigstein:2023yzv,Pannullo:2023cat}),
 while at $d=2$ it vanishes when the cutoff is fully removed \cite{Buballa:2020nsi,Pannullo:2021edr,Pannullo:2023one,Winstel:2022jkk}.
This also happens for noninteger 2<d<3 \cite{Koenigstein:2023yzv,Pannullo:2023cat}.
 Finally, for the physical case $d=3$, 
	it is no longer possible to renormalise the theory and the location of the inhomogeneous phase in the temperature--chemical
	potential plane as well as its very existence depend on the scheme and scale of the regularisation
 \cite{Pannullo:2022eqh}. 
 It therefore 
	becomes difficult to disentangle effects that might be related to a potential physical content of the regularisation 
	procedure from simple cut-off artefacts.

	This problem is absent in QCD. In turn, however, one has to deal with all the complications of a full-fledged non-perturbative
	gauge theory. Since the large chemical potential region of the QCD phase diagram is not (yet) accessible by lattice QCD (due to
	the sign-problem, see \cite{Berger:2019odf,Nagata:2021ugx} for recent reviews), effective theories and functional methods are
	methods of choice. The problem is very difficult, however, and only very few explicit studies are available. Signals for 
	instabilities to so-called moat-regimes around the QCD critical point have been found in Ref.~\cite{Fu:2019hdw}. The only 
	explicit microscopic study of inhomogeneous phases in the small temperature and large chemical potential regime so far has been 
	reported in Ref.~\cite{Muller:2013tya}. Using a coupled set of Dyson-Schwinger equations (DSEs) for the quark and gluon propagators
	in the massless $N_f=2$ limit an inhomogeneous chiral density wave has been found to be favoured over 
 homogeneous phases in a certain region of the phase diagram.
 In general, explicit studies of this 
	type need to assume a specific form of inhomogeneity and may therefore be limited to simple cases.
	
    An alternative method 
    is the stability analysis of the homogeneous ground state 
    with respect to inhomogeneous fluctuations. 
    In contrast to explicit calculations \cite{Muller:2013tya} this type of analysis does not rely on 
    specific forms of inhomogeneity and is therefore not restricted to simple cases.
    It has been successfully applied to the Nambu-Jona-Lasinio (NJL) model    \cite{Buballa:2018hux,Carignano:2019ivp,Pannullo:2022eqh}, GN-model \cite{Thies:2003kk,Buballa:2020nsi,Koenigstein:2021llr} and the 
    quark-meson model \cite{Buballa:2020xaa}. 
    In Ref.~\cite{Motta:2023pks} we have generalized the method to be applicable to QCD.
    This method is valid for any truncations of the two-particle irreducible (2PI) effective action and is therefore suitable 
    for simple models as well as for full QCD.  
    In Ref.~\cite{Motta:2023pks}, we have tested it by applying it to a simple truncation of QCD and investigating 
    the stability of the chirally symmetric phase against {\it homogeneous} chiral-symmetry breaking fluctuations. 

    In this paper, we perform an exploratory stability analysis of QCD with respect to {\it inhomogeneous} fluctuations in a simple rainbow-ladder type approximation of the quark-gluon interaction using the technique described in \cite{Motta:2023pks}. Our study serves to provide strategies for the identification of the most effective test functions for the analysis and therefore provides important steps towards a more refined analysis with more realistic truncation schemes. 
    The paper is organised as follows. 
    In section \ref{sec:stabilityanalysis} we briefly review the stability analysis method introduced in Ref.~\cite{Motta:2023pks} and then discuss in details our strategy for applications to	potential inhomogeneous phases.      
    In section \ref{sec:2pi} we discuss our truncation of the 2PI effective action and the resulting DSE theoretical framework. We present cross-checks and results for instabilities towards inhomogeneous phases 
    in section \ref{sec:results} and conclude in section \ref{sec:conclusions}.
    
	\section{2PI Stability Analysis}\label{sec:stabilityanalysis}
	
	%
	The 2PI effective action of QCD in the absence of background fields can be denoted as follows\footnote{In this publication we will use the capitalised ``Tr'' notation for functional traces, i.e., the combination of integration over all continuous quantities (e.g., momentum or coordinate space) together with the trace over discrete Dirac, colour, and flavour indices. Whenever the continuous or discrete indices are omitted, the standard cyclic pattern is assumed to be understood, e.g.,
	$$
	\text{Tr}[S_0^{-1}S] =
	\int_{k_1,k_2}\sum_{\tiny
		\begin{matrix}
		D_1,D_2,\\ c_1,c_2,\\f_1,f_2
		\end{matrix}
	}
	\big(S_0^{-1}(k_1,k_2)\big)_{D_1,D_2,c_1,c_2,f_1,f_2}
	\big(S(k_2,k_1)\big)_{D_2,D_1,c_2,c_1,f_2,f_1}
	$$
	where $D_i$, $c_i$ and $f_i$ are Dirac, colour and flavour indices, respectively. Note also that we have changed the convention concerning $\Gamma$ compared to our previous work \cite{Motta:2023pks}.
	}
\begin{equation}\label{eq:2PI}
	\begin{aligned}
		\Gamma[S,D,\Delta]&=\Tr\log\left[S^{-1}S_0\right] + \Tr\left[S_0^{-1}S\right] && \text{\color{darkgray}$\rightarrow$ Quark ``kinetic'' part}
		\\ &-\frac{1}{2}\Big(\Tr\log\left[D^{-1}D_0\right] 
		+ \Tr\left[D_0^{-1}D\right]\Big) && \text{\color{darkgray}$\rightarrow$ Gluon ``kinetic'' part}
		\\ &+\Tr\log\left[\Delta^{-1}\Delta_0\right] + \Tr\left[\Delta_0^{-1}\Delta\right] && \text{\color{darkgray}$\rightarrow$ Ghost ``kinetic'' part}
		\\ &+\Phi[S,D,\Delta] && \text{\color{darkgray}$\rightarrow$ Interaction part}
	\end{aligned}
\end{equation}
	 Here the dressed quark propagator is denoted by $S$, the dressed gluon propagator by $D$ and the Faddeev-Popov ghost propagator
	 by $\Delta$. The corresponding bare quantities carry a subscript zero and we have resorted to a symbolic notation without 
	 Lorentz, colour and flavour indices. The effective action is conveniently split into a kinetic part, given explicitly, and an
	 interaction part denoted by $\Phi$. For practical calculations, the kinetic and interaction part of the effective action are subject to truncations. In later sections of this work we will indeed resort to a (simple) truncation of the effective action to demonstrate the feasibility of our approach. We would like to stress, however, that all results of this section are completely general and can be used in future work to study improved truncations closer to QCD. 

    For the following discussion, we may focus on the quark propagator. 
    Suppose $\bar{S}$ is a stationary point of the 2PI effective action.
    In order to analyse whether this corresponds to a stable solution we consider small 
    fluctuations $\epsilon_S$ around this point,  $S = \bar{S} + \epsilon_S$, 
    calculate the free-energy functional $\Omega[S]=-\frac{T}{V} \Gamma[S]$ and perform a functional Taylor expansion around $\bar{S}$
	\begin{equation}\label{eq:stabilityI}
	\Omega[S] = 
	\Omega[\bar{S}]
	+
	\Tr\Bigg[\left.\frac{\delta \Omega}{\delta S}\right|_{S=\bar{S}} \epsilon_S \Bigg]
	+
	\Tr\Bigg[\left.\frac{\delta^2 \Omega}{\delta S\delta S}\right|_{S=\bar{S}}  \epsilon_S\epsilon_S \Bigg]
	+ \cdots .
	\end{equation}
	The terms in the r.h.s. of the equation above we call respectively $\Omega^{(0)}$, $\Omega^{(1)}$, $\Omega^{(2)}$, etc.
	At the stationary point $\Omega^{(1)}$ is zero by the definition. We then take the leading non-trival order $\Omega^{(2)}$
	and analyse its sign. If the sign is \textit{negative}, we can conclude that the homogeneous ground state is not the lowest 
	energy state and thus unstable. This is the (in-)stability criterion.
	
	Most generally, the quark propagator is defined as
	$$
	S(x,y) = \avg{\mathcal{T}\psi(x)\bar\psi(y)},
	$$
	that is, the time-ordered product of fermionic field operators in different space-time coordinates.
	In the homogenous case, as a consequence of translational invariance, the propagator depends only on the difference of the coordinates,
	\begin{equation}\label{prop_bar_def_config}
	S(x,y) \equiv {S}(x-y).
	\end{equation}
 	This is equivalent to saying that momentum is conserved, i.e., the momentum-space structure of the propagator is given by\footnote{For notational convenience we refrain from introducing different symbols for the general propagator and its diagonal part in the homogeneous case.  In the following they can always be distinguished by the number of their momentum arguments (two or one, respectively).} 
	\begin{equation}\label{prop_bar_def}
	S(k_1,k_2) \equiv {S}(k_1)\delta(k_1-k_2),  
    \end{equation}
    where $k=(\vec{k},\omega_k)$ are Euclidean four-momenta with fermionic Matsubara frequencies $\omega_k=(2l_q+1)\pi T$ and three-momenta $\vec{k}$.  
    Naturally, if the propagator is \textit{not} diagonal in momentum space, i.e., if off-diagonal terms of $S(k_1,k_2)$ are non-zero, this is a manifestation of translational symmetry breaking, corresponding to an inhomogeneous state. 

    The Dirac structure of the homogeneous propagator is given by
    \begin{equation}\label{eq:quark}
	S(k) = \frac{1}{i \slashed{\vec{k}}A(k) +B(k) +i(\omega_k+i\mu)\gamma_4C(k)},
	\end{equation}
    with dressing functions $A$, $B$ and $C$.
    Chiral symmetry is broken by a non zero value of the quark mass. As can be seen from Eq.~(\ref{eq:quark}), the Dirac-scalar component of the propagator, the $B$ dressing function, acts as a proxy for the dynamical mass of the quark\footnote{The precise definition of the dynamical mass being $M(k) = \frac{B(k)}{C(k)}$} and it can thus be used as an order parameter for the breaking of chiral symmetry. It also gives rise to a non-zero quark-condensate $\avg{\bar\psi\psi}\propto\Tr[S]$.
	Inhomogeneous chiral symmetry breaking can be defined by the combined breaking of translational symmetry and chiral symmetry. Non-zero, Dirac-scalar, off-diagonal terms of the momentum-space propagator break chiral symmetry inhomogeneously.

    In Ref.~\cite{Motta:2023pks} we tested the method outlined above by applying it
    to a stationary point of the effective action that is both chiral-symmetric and translational-invariant, i.e.,
	\begin{equation}\label{eq:quarkexample1}
	\bar S(k) = \frac{1}{i \slashed{\vec{k}}A(k) +i(\omega_k+i\mu)\gamma_4C(k)}
	= \frac{-i \slashed{\vec{k}}A(k) -i(\omega_k+i\mu)\gamma_4C(k)}{ {\vec{k}}^2A(k)^2+(\omega_k+i\mu)^2C(k)^2}.
	\end{equation}
    Then we analysed whether or not this solution is stable with respect to \textit{translational-invariant} chiral symmetry breaking. 
    This was done by adding a \textit{small} effective mass to this solution,
    or more precisely a small $B$ function, which we call $\epsilon_m(k)$.  
    Accordingly, the propagator receives an extra term\footnote{Note that in theory a $B$ function would also appear in the denominator of a chirally broken propagator. However, since $\epsilon_m$ is assumed to be small, we can neglect $\epsilon_m^2$ terms.}
	\begin{equation}\label{eq:quarkexample2}
	\epsilon_S(k)  = \frac{\epsilon_m(k)}{ {\vec{k}}^2A(k)^2+(\omega_k+i\mu)^2C(k)^2}.
	\end{equation}
If such an addition to the propagator causes the system to lower its free-energy, then the chiral-symmetric system is \textit{unstable} with respect to homogeneous chiral symmetry breaking. 
Applying this ``chiral test'' to an example case, we found in Ref.~\cite{Motta:2023pks} that the resulting second-order chiral phase boundary agrees
well with the one obtained from directly solving the homogeneous DSE.

    Our goal in this paper, however, is to investigate also the breaking of translational symmetry. 
    As seen above, we must then consider additions to the propagator which are non-diagonal in momentum space, i.e., we take the full propagator to be
	\begin{equation}
	S(k_1,k_2) = \bar{S}(k_1)\delta(k_1-k_2) + \epsilon_S(k_1,k_2),
	\end{equation}
    where, as before, the expansion point $\bar{S}(k_1,k_2) \equiv \bar{S}(k_1)\delta(k_1-k_2)$
    corresponds to a homogeneous stationary point of the effective action. 
    Moreover, throughout this paper we work in the chiral limit and    
    restrict ourselves to chirally symmetric expansion points,
    i.e., $\bar{S}(k)$ is given by Eq.~(\ref{eq:quarkexample1}).
	
	An alternative and useful re-formulation of the stability criterion arises by introducing an inhomogeneous contribution to the quark
	\textit{self-energy} rather than the propagator. Let $\epsilon_\Sigma(k_1,k_2)$ be an inhomogeneous perturbation to the 
	self-energy. We can exactly relate one to the other via the Dyson series, resulting in the following expression\footnote{For 
	more information see Ref.~\cite{Motta:2023pks}. Note also that in our previous work, the fluctuations
	$\epsilon_S$ and $\epsilon_\Sigma$ were denoted by a different notation, $\delta S$ and $\delta \Sigma$.}
	\begin{equation} \label{eq:stabilityII}
	\epsilon_S(k_1,k_2)=\bar{S}(k_1)\epsilon_\Sigma(k_1,k_2)\bar{S}(k_2),
	\end{equation}
	and its inverse.
	This is nothing but a physically motivated change of variables, from $\Omega^{(2)}[\epsilon_S]$ to $\Omega^{(2)}[\epsilon_\Sigma]$.
	In the next section, we will discuss both choices further by example of our simple rainbow-ladder truncation. While	at first sight
	the first formulation may seem more natural, since $\Gamma$ is a functional of the propagator, we will then realise that the
	second formulation is technically and physically advantageous.
	
	Note, that a stability analysis in simple models like the NJL-model can be performed without explicit expressions for the
	perturbations $\epsilon_S(k_1,k_2)$ or $\epsilon_\Sigma(k_1,k_2)$. This is due to the self-energy being local in these models. Then $\epsilon_\Sigma$ is a function only of $k_1-k_2$ and can be factored out, as discussed in detail in Ref.~\cite{Motta:2023pks}. 
	Since this is not the case for QCD, we discuss explicit test-functions for the perturbations in the next two subsections.
	
	\subsection{The test function - general form}
	
	Before we discuss details of our explicit test-function $\epsilon_\Sigma(k_1,k_2)$ we would like to stress that this is 
	\textit{not} an ansatz for the true self-consistent solution of the inhomogeneous propagator, if one exists. The test-function 
	is simply a means to trigger an instability. In Ref.~\cite{Motta:2023pks} we explicitly showed that the instability of the chiral Wigner-Weyl (WW) solution at low $T$ and $\mu$
    with respect to homogeneous chiral-symmetry breaking fluctuations 
    can be triggered by infinitely many test-functions 
	that may or may not have much similarity with the actual shape of the homogeneously broken Nambu-Goldstone (NG) solution. 
	The same applies to instabilities with respect to inhomogeneous fluctuations.
	  
	Nevertheless, it seems reasonable to take a test-function that we know might be similar in shape to a self-consistent 
	inhomogeneous solution. For the perturbation of the self-energy, we therefore base our analysis on a Dirac-scalar 
	function of the form
	\begin{equation}\label{test-functionSig}
	\epsilon_\Sigma(k_1,k_2) =\left(\frac{\epsilon_m(k_1)+\epsilon_m(k_2)}{2}\right)F(k_1-k_2),
	\end{equation}
	where the $\epsilon_m(k)$ can be related to a test-function used to probe homogeneous chiral symmetry breaking, see Eq.~(\ref{eq:quarkexample2}). These functions $\epsilon_S$, $\epsilon_m$ and $\epsilon_\Sigma$ all have certain properties that have to be maintained. E.g., the full propagator's so-called adjoint relation
	\begin{equation}\label{adjoint}
	S(\omega_1,\vec{k}_1,\omega_2,\vec{k}_2)^\dagger = \gamma_4 S(-\omega_2,\vec{k}_2,-\omega_1,\vec{k}_1) \gamma_4
	\end{equation}
	forces, amongst other things, that the $F$ function in Eq.~(\ref{test-functionSig}) be such that $F(-k)=F(k)^\star$, which is equivalent to saying that in coordinate space $F$ is a real-valued function.
	
	This choice of overall structure follows two main reasons:
    \begin{enumerate}
	\item The choice is much more general but also includes the special case $F(k_1-k_2)=\delta(k_1-k_2+Q)$,
    which correponds to the momentum structure of the inhomogeneous chiral density wave\footnote{The 
    self-consistent solution in Ref.~\cite{Muller:2013tya} contains additional structures in Dirac and flavour space. However, as argued above, for the test-function it is sufficient to have some component in an unstable direction.}
    found in Ref.~\cite{Muller:2013tya} to be energetically favoured over the chiral symmetric phase in
	a certain region of the phase diagram. As we will see below, in our case the $F$ function even factors out and does not influence 
	the analysis. Thus the main point of Eq.~(\ref{test-functionSig}) is the specific Dirac and momentum structure, not the
	precise form of $F$. 
	
	\item We can use the limit $k_1=k_2\equiv k$ to test whether or not the setup reproduces the instability of the homogeneous chiral 
	solution with respect to homogeneous chiral symmetry breaking. In this case the propagator's test function
	\begin{equation}\label{test-functionS}
		\epsilon_S(k_1,k_2) = \bar{S}(k_1) \left(\frac{\epsilon_m(k_1)+\epsilon_m(k_2)}{2}\right)\bar{S}(k_2)\times F(k_1-k_2)
	\end{equation}
	becomes
	\begin{equation}\label{test-functionShomog}
		\epsilon_S(k,k) = -\frac{\epsilon_m(k)}{{\vec{k}^2A(k)^2+(\omega_k+i\mu)^2C(k)^2}} \times F(0),
	\end{equation}
	which, apart from the irrelevant factor $-F(0)$, is equal to Eq.~(\ref{eq:quarkexample2}), i.e., it corresponds to adding a small dynamical mass 
    $\epsilon_m(k)$ to the quarks, breaking 
	chiral symmetry homogeneously.	Since fixing the parameters of the test-function is far from trivial (see next section), 
	it is helpful to have tests one can perform to assure the consistency of this approach. 
    \end{enumerate}

	\subsection{The test function - matching real and imaginary parts}
	
	At finite chemical potential, all dressing functions of the quark propagator are complex in general. This complicates the 
	analysis greatly since every stationary point becomes a saddle point in the complex plane. Consequently, our test functions
	for the	perturbations also need to be complex and one can find instabilities everywhere by picking the wrong imaginary part 
	for the test function (including zero). This has been noted also in Refs.~\cite{Haensch:2023sig} and \cite{Motta:2023pks}.
	
	One natural solution to this problem is to simply take the direction of steepest ascent of the free-energy, i.e. for every 
	real part of $\epsilon_\Sigma$ we test, the imaginary part has to be such that it maximises $\Omega^{(2)}$. Naturally, since 
	we want to investigate all possible instabilities, we would like to find the real part that minimises $\Omega^{(2)}$. 
	This will amount to finding a saddle point of $\Omega^{(2)}[\epsilon_\Sigma]$ in the complex plane. That this is even more
	than a convenient construction principle has been pointed out in Refs.~\cite{winstel2024spatially,Haensch:2023sig} in the context of a mean-field 
	1PI effective action: In complex actions one can always	redefine the integration path of the generating	functional to be 
	along the Lefschetz Thimble (LT). If the fields live on the thimble, so must the perturbations. Therefore, taking the 
	direction of steepest ascent of the free energy\footnote{Note that the direction of steepest ascent of the free energy
	equals the steepest descent of pressure, which, due to it being the log of the generating functional, equals the steepest 
	descent of $\mathcal{Z}$ which equals the steepest ascent of the classical action.} might not only be convenient, but the 
    only correct direction to be taken since the LT goes along the steepest ascent (Ref.~\cite{Mori_2018} calculates the LT explicit in an NJL model with vector repulsion). Although in our case (2PI beyond mean field)
    the connection between thimble and stability analysis is not as apparent as in Ref.~\cite{Haensch:2023sig}, we regard their
    analysis as an additional motivation that we should indeed fix whatever parameters are left of our test function to 
    find a complex saddle point of the stability condition. 
     
	For our stability analysis we take the test-function to be of the form
	\begin{equation}\label{epsm}
		\epsilon_m(\omega,\vec{k}) = 
		e^{-\left(\frac{\omega^2}{L_1^2}+\frac{\vec k^2}{L_2^2}\right)}
		+ i L_3 \frac{\omega}{\omega_0} e^{-\left(\frac{\omega^2}{L_4^2}+\frac{\vec k^2}{L_5^2}\right)}
	\end{equation}
	where the parameters $L_n$ will be fixed according to the saddle point condition, and where $\omega_0$ is the lowest Matsubara
	frequency at any given temperature. Note that the linearity of the imaginary part with respect to $\omega$ is chosen to ensure
	$\epsilon_m(-\omega,\vec{k})=\epsilon_m(\omega,\vec{k})^\star$ which follows from the way the dressing functions transform over
	time-reversal.
 
 Eq.~(\ref{epsm}) is a generalisation of the form used in Ref.~\cite{Motta:2023pks} for instabilities of the 
	chirally symmetric phase towards the homogeneously broken one. Since inhomogeneous phases are expected to appear in lower rather than high temperatures, in this work we have to try and reach further down the temperature axis of the phase diagram. As it turns out, here we need more 
	parameters than in \cite{Motta:2023pks} in particular in the imaginary part to accommodate for the more complex physics at the smaller temperatures. We have verified, however, that increasing the number of parameters does not change our previous results of
	Ref.~\cite{Motta:2023pks}. All of the parameters will be fixed by the saddle point condition as further detailed below.

	\section{2-Loop 2PI Effective Action \& Dyson-Schwinger Equations}\label{sec:2pi}
	
	Up to now, all our considerations are general in the sense that they are independent of the specific truncation of the
	effective action. Thus, they support our ultimate goal to apply the formalism to approximations of QCD that can be 
	systematically improved towards the full theory. One way to do so is by using loop expansions of 2PI (or, more general, nPI)
	effective actions \cite{berges2004n}. As a first step in this direction we employ a rainbow-ladder type approximation 
	that will be detailed in this section.
	
	We start from the 2PI effective action, Eq.~(\ref{eq:2PI}) and model the gauge sector by taking an analytic function that 
	reproduces known global properties of the gluon propagator and the quark-gluon vertex. We only take into account the matter part of the effective action, neglecting the ghost and
	gluon kinetic terms.	
    \begin{equation}\label{CJT}
	\Gamma[S]=\Tr\log\left[S^{-1}S_0\right] + \Tr\left[S_0^{-1}S\right] +\Phi[S].
    \end{equation}
    We furthermore truncate the interaction part at two-loop order
    \begin{equation}\label{2pi}
	\Phi_\text{2Loop}^{qg}[S]=\frac{1}{2}
	\raisebox{-0.8cm}{
		\begin{tikzpicture}
			\begin{feynman}
				\vertex (a);
				\vertex [right=of a] (b);
				\vertex [right=of a] (b);
				\vertex [right=0.75cm of a] (l);
				\diagram*{
					(a) -- [fermion3,half left] (b);
					(b) -- [boson] (a);
					(b) -- [fermion3,half left] (a);
				};\draw (l) node [gray, dot];
			\end{feynman}
		\end{tikzpicture}
	}
	=\frac{1}{2}\text{Tr}
	\left[g^2
	Z_{1F} \gamma_\mu t^a  S(k_1,k_2) \gamma_\nu t^b \Gamma_{qg}(q)
	S(k_2-q,k_1-q) D_{\mu\nu}^{ab}(q)
	\right],
    \end{equation}
    with one bare quark-gluon vertex 
    (with bare coupling $g$ and colour matrices $t^a$)    
    renormalized by the vertex renormalization factor $Z_{1F}$, 
    one dressed vertex carrying the tree-level tensor structure only and dressed with a function $\Gamma_{qg}(q)$ that 
    depends only on the gluon momentum. 
    The black dot symbolises the fully dressed quark propagator. The gray dot on the gluon propagator represents
    the combination of the vertex dressing $\Gamma_{qg}(q)$ and the dressing of the gluon propagator that will be 
    represented by an analytic model detailed below.

    In models of this type, the term whose sign constitutes the stability condition
    is given by
    \begin{equation}\label{SC1}
	  \begin{aligned}
		\Omega^{(2)}[\epsilon_S]=
		&-\Tr\left[
		\bar{S}^{-1}(k_1)
		\epsilon_S(k_1,k_2)
		\bar{S}^{-1}(k_2)
		\epsilon_S(k_2,k_1)
		\right]\\
		&-g^2
	Z_{1F}\text{Tr}
		\left[
		\gamma_\mu t^a  \epsilon_S(k_1,k_2) \gamma_\nu t^b
		\epsilon_S(k_2-q,k_1-q) D_{\mu\nu}^{ab}(q)\Gamma_{qg}(q)
		\right]\,,
	  \end{aligned}
    \end{equation}
    if we expand the free energy in perturbations of the the propagator as in Eq.~(\ref{eq:stabilityI}).
    Employing Eq.~(\ref{eq:stabilityII}) to write this in terms of self-energy perturbations, this becomes
   \begin{equation}\label{SC2}
	\begin{aligned}
			\Omega^{(2)}[\epsilon_\Sigma]=
			&-\Tr\left[
			\bar{S}(k_1)
			\epsilon_\Sigma(k_1,k_2)
			\bar{S}(k_2)
			\epsilon_\Sigma(k_2,k_1)
			\right]\\
			&-g^2
	Z_{1F}\text{Tr}
			\left[
			\gamma_\mu t^a  \bar{S}(k_1)\epsilon_\Sigma(k_1,k_2)\bar{S}(k_2) \gamma_\nu t^b
			\bar{S}(k_2-q)\epsilon_\Sigma(k_2-q,k_1-q)\bar{S}(k_1-q) D_{\mu\nu}^{ab}(q)\Gamma_{qg}(q)
			\right].
		\end{aligned}
   \end{equation}
   Inserting Eq.~(\ref{test-functionSig}) for $\epsilon_\Sigma$ we now can explicitly verify 
   a statement that has been mentioned already in the previous section: 
   We \textit{can} be agnostic with respect to the functional form of $F(k_1-k_2)$ in $\epsilon_\Sigma$. In Eq.~(\ref{SC2}) 
   the momentum dependence on the test functions will cause $F(k_1-k_2)$ and $F(k_2-k_1)$ 
   to appear in both terms. 
   As pointed out before, these functions are real in coordinate space and thus have the property
   $F(k) = F(-k)^\star$ in momentum space.   
   We therefore arrive at\footnote{Throughout this paper we will use the notation
   	$\SumInt_p = T\sum_{\omega_n}\int_{\vec{p}}= T\sum_{\omega_n}\int\frac{d^3p}{(2\pi)^3}$ for integration over the 3-momentum domain and sum over the relevant Matsubara frequencies. 
    The lower case ``tr'' means trace over discrete indices such as Dirac, colour, flavour, and so on, as opposed to the capitalised ``Tr'' notation which involves also a trace over continuum variables such as momentum or coordinate space.}
   \begin{equation}\label{SC3}
	 \begin{aligned}
		\Omega^{(2)}[\epsilon_\Sigma]=
		&-\SumInt_{k_1 k_2} |F(k_1-k_2)|^2\times\text{tr}\Bigg[
		\bar{S}(k_1)
		\left(\frac{\epsilon_m(k_1)+\epsilon_m(k_2)}{2}\right)
		\bar{S}(k_2)
		\left(\frac{\epsilon_m(k_2)+\epsilon_m(k_1)}{2}\right)
		\Bigg]\\
		&
		\begin{aligned}-g^2
	Z_{1F}\SumInt_{k_1 k_2 q} |F(k_1-k_2)|^2\times
			\text{tr}\Bigg[&\gamma_\mu t^a  \bar{S}(k_1)
			\left(\frac{\epsilon_m(k_1)+\epsilon_m(k_2)}{2}\right)
			\bar{S}(k_2)\\
			&\gamma_\nu t^b
			\bar{S}(k_2-q)
			\left(\frac{\epsilon_m(k_2-q)+\epsilon_m(k_1-q)}{2}\right)
			\bar{S}(k_1-q) D_{\mu\nu}^{ab}(q)\Gamma_{qg}(q)	\Bigg]
		\end{aligned}
 	 \end{aligned}
    \end{equation}
    or, after substituting $k_1 = k+d$, $k_2 = k-d$ and $q = k-l$, 
    \begin{equation}\label{SC3}
    \begin{aligned}
    \Omega^{(2)}[\epsilon_\Sigma]=
    &-\frac{1}{4}\SumInt_{kd} |F(2d)|^2\times\text{tr}\Bigg[
    \bar{S}(k+d)
    \left({\epsilon_m(k+d)+\epsilon_m(k-d)}\right)
    \bar{S}(k-d)
    \left({\epsilon_m(k-d)+\epsilon_m(k+d)}\right)
    \Bigg]\\
    &
    \begin{aligned}-\frac{g^2
	Z_{1F}}{4}\SumInt_{kdl} |F(2d)|^2\times
    \text{tr}\Bigg[&\gamma_\mu t^a  \bar{S}(k+d)
    \left({\epsilon_m(k+d)+\epsilon_m(k-d)}\right)
    \bar{S}(k-d)\\
    &\gamma_\nu t^b
    \bar{S}(l-d)
    \left({\epsilon_m(l-d)+\epsilon_m(l+d)}\right)
    \bar{S}(l+d) D_{\mu\nu}^{ab}(k-l)\Gamma_{qg}(k-l)\Bigg],
    \end{aligned}
    \end{aligned}
    \end{equation}
    which naturally can be written as
    \begin{equation}\label{eq:Omt1}
    	\Omega^{(2)} = \SumInt_d |F(2d)|^2 \times  \tilde \Omega^{(2)}(d).
    \end{equation}
    Consequently, the function 
     \begin{equation}\label{eq:Omt2}
    \begin{aligned}
    \tilde \Omega^{(2)}(d)=
    &-\frac{1}{4}\SumInt_{k} \text{tr}\Bigg[
    \bar{S}(k+d)
    \left({\epsilon_m(k+d)+\epsilon_m(k-d)}\right)
    \bar{S}(k-d)
    \left({\epsilon_m(k-d)+\epsilon_m(k+d)}\right)
    \Bigg]\\
    &
    \begin{aligned}-\frac{g^2
	Z_{1F}}{4}\SumInt_{kl}
    \text{tr}\Bigg[&\gamma_\mu t^a  \bar{S}(k+d)
    \left({\epsilon_m(k+d)+\epsilon_m(k-d)}\right)
    \bar{S}(k-d)\\
    &\gamma_\nu t^b
    \bar{S}(l-d)
    \left({\epsilon_m(l-d)+\epsilon_m(l+d)}\right)
    \bar{S}(l+d) D_{\mu\nu}^{ab}(k-l)\Gamma_{qg}(k-l)	\Bigg],
    \end{aligned}
    \end{aligned}
    \end{equation}       
    can itself be used as a stability condition: If $\tilde \Omega^{(2)}$ is 
    ever negative for \textit{any} region of the momentum domain, one can conceive of an $F$ function which peaks in the same 
    region and causes $\Omega^{(2)}$ to become negative.
    
    As per usual, we consider phases which are inhomogeneous in space, but static in time. This amounts to imposing one condition on the $F$ function which is
    \begin{equation}
    F(k_1-k_2)_\text{static} = F(\vec{k}_1 - \vec{k}_2)\delta_{\omega_1 , \omega_2},
    \end{equation}
	where $\omega_1$ and $\omega_2$ are the Matsubara frequencies of the Euclidian 4-momenta ${k_1} = (\vec{k}_1,\omega_1)$, and similarly for $k_2$. 
    As a consequence, Eq.~(\ref{eq:Omt1}) becomes
	\begin{equation}
	\Omega^{(2)} = \int_{\vec{d}}|F(2\vec d)|^2 \times  \tilde \Omega^{(2)}(d),
	\end{equation}
    where $\tilde\Omega^{(2)}(d)$ is given by Eq.~(\ref{eq:Omt2}) evaluated for  
    $d=(\vec{d},0)$. In fact, since the homogeneous propagators and our test functions are isotropic, $\tilde\Omega^{(2)}$ only depends on the modulus $|\vec d|$, 
    which we will also call just $d$.

 
   The explicit structure of the chiral-symmetric 
   homogeneous quark propagator $\bar{S}$ 
   was already given in Eq.~(\ref{eq:quarkexample1}).     
   The gluon propagator $D_{\mu\nu}$ in Landau gauge is given by 
    \begin{align}\label{eq:qProp}
	D_{\mu\nu}^{ab}(q) &= \Bigg(P_{\mu\nu}^{T}(q) \,\frac{Z_{T}(q)}{q^2} + P_{\mu\nu}^{L}(q) \,\frac{Z_{L}(q)}{q^2} \Bigg)\delta^{ab} \,,
    \end{align}
	with momentum $q=(\vec{q},\omega_q)$ and $\omega_q=2l_q \pi T$ are the bosonic Matsubara frequencies.
	The projectors $P_{\mu\nu}^{{T},{L}}$ are transverse (${T}$) and longitudinal (${L}$) with respect
	to the heat-bath vector aligned in four-direction and given by
	\begin{align}\label{eq:projTL}
	P_{\mu\nu}^{T} &= \left(1-\delta_{\mu 4}\right)\left(1-\delta_{\nu 4}\right)\left(\delta_{\mu\nu}-\frac{q_\mu q_\nu}{\vec{q}^{\,2}}\right), \hspace*{2cm}
	P_{\mu\nu}^{L} = P_{\mu\nu} - P_{\mu\nu}^{T} \,,
	\end{align}
	where $P_{\mu\nu} = \delta_{\mu\nu} - q_\mu\, q_\nu/q^2$ is the covariant transverse projector.

    We then combine the dressings of the gluon, the dressing of the (tree-level structure of the) quark-gluon vertex and
    the coupling $g^2$ together with renormalization factors\footnote{In the end only one renormalization constant is needed in this truncation and it is fixed by the condition $$S^{-1}(\zeta)=(ip+m)\vert_{p=\zeta},$$ at a scale $\zeta=19$GeV, which amounts to 
    $$Z_2=(1+\Sigma_A(\zeta))^{-1},$$ where $\Sigma_A$ is the Dirac vector component of the quark self-energy. The renormalization constant is obtained in vacuum and maintained fixed for finite $T$ and $\mu$.} of the vertex, $Z_{1F}$, and the quark, $Z_2$ (which in this truncation is equal to $Z_{1F}$), 
    into the 
    following expression for a renormalization group invariant effective running coupling
    \begin{equation}
    	\alpha_{T,L}(q) = \frac{g^2}{4 \pi} \frac{Z_{1F}}{Z_2^2} \Gamma_{qg}(q) Z_{T,L}(q) 
    \end{equation}
    In this paper we discuss results using three different models for this effective running coupling, namely:
    \begin{itemize}
	\item The Maris-Tandy model (MT) \cite{Maris:1999nt}
	\begin{equation}
		\begin{aligned}
			\alpha_\text{MT}(q^2) = 
			\pi \frac{\eta^7}{\Lambda_\text{MT}^4} q^4 e^{-\eta^2\,q^2/\Lambda_\text{MT}^2} + 
			\frac{4\pi^2\gamma_m}{(1/2)\log(\tau+(1+q^2/\Lambda_\text{QCD}^2)^2)}
			\Big({1-e^{-q^2/4m_t^2}}\Big) \,,
		\end{aligned}
	\end{equation}
	which was introduced in the context of hadron spectroscopy, see e.g. \cite{Maris:2003vk,Eichmann:2016yit} for a discussion of corresponding results. 
    The second term on the right-hand side corresponds to the perturbative running in the ultraviolet,
    while the first term models the infra-red behaviour.
 Although the model does not deliver realistic values for the QCD end point (see \cite{Fischer:2018sdj} 
	for an overview) it serves well as a starting point due to its simplicity. Here we use the usual 
 parameters of $\eta = 1.8$, $\Lambda_\text{MT}=720$ MeV, $\gamma_m=12/(33-2N_F)$, $N_F=4$, $m_t=500$ MeV, $\Lambda_\text{QCD}=234$ MeV and $\tau = e^2-1$.
	
	\item The infra-red Maris-Tandy (MTIR) model. Since the ultraviolet log-tail of the MT-model often turns out to be quantitatively 
	irrelevant it may be neglected to even further simplify the model. This is technically advantageous, as it makes the model
	super-renormalisable. Since it is interesting to check the effect of the UV part of the coupling for inhomogeneous 
	phases we also explore the choice
	\begin{equation}
		\begin{aligned}
			\alpha_\text{MTIR}(q^2) = 
			\pi \frac{\eta^7}{\Lambda_\text{MT}^4} q^4 e^{-\eta^2\,q^2/\Lambda_\text{MT}^2} \,.
		\end{aligned}
	\end{equation}
	
	\item The infra-red Qin-Chang (QCIR) model \cite{Qin:2011dd} is an alternative choice for the low momentum part of the effective
	running coupling
	\begin{equation}
		\begin{aligned}
			\alpha_\text{QCIR}(q^2) = 
			\frac{2\pi}{\omega_\text{QC}^4} D q^2 e^{-q^2/\omega_\text{QC}^2},,
		\end{aligned}
	\end{equation}
    which differs from MTIR essentially by the missing factor of $q^2$, which makes it,
    when combined with the factor $1/q^2$ from Eq.~(\ref{eq:qProp}),
    infra-red finite instead of vanishing.
    There is also a variant of this model with UV logarithmic tail, which we do not consider in this work. 
	We chose the parameters $\omega_\text{QC}=600$~MeV and $D=1$~GeV$^2$. 
    \end{itemize}
    \begin{figure}
	\centering
	\includegraphics[width=0.5\linewidth]{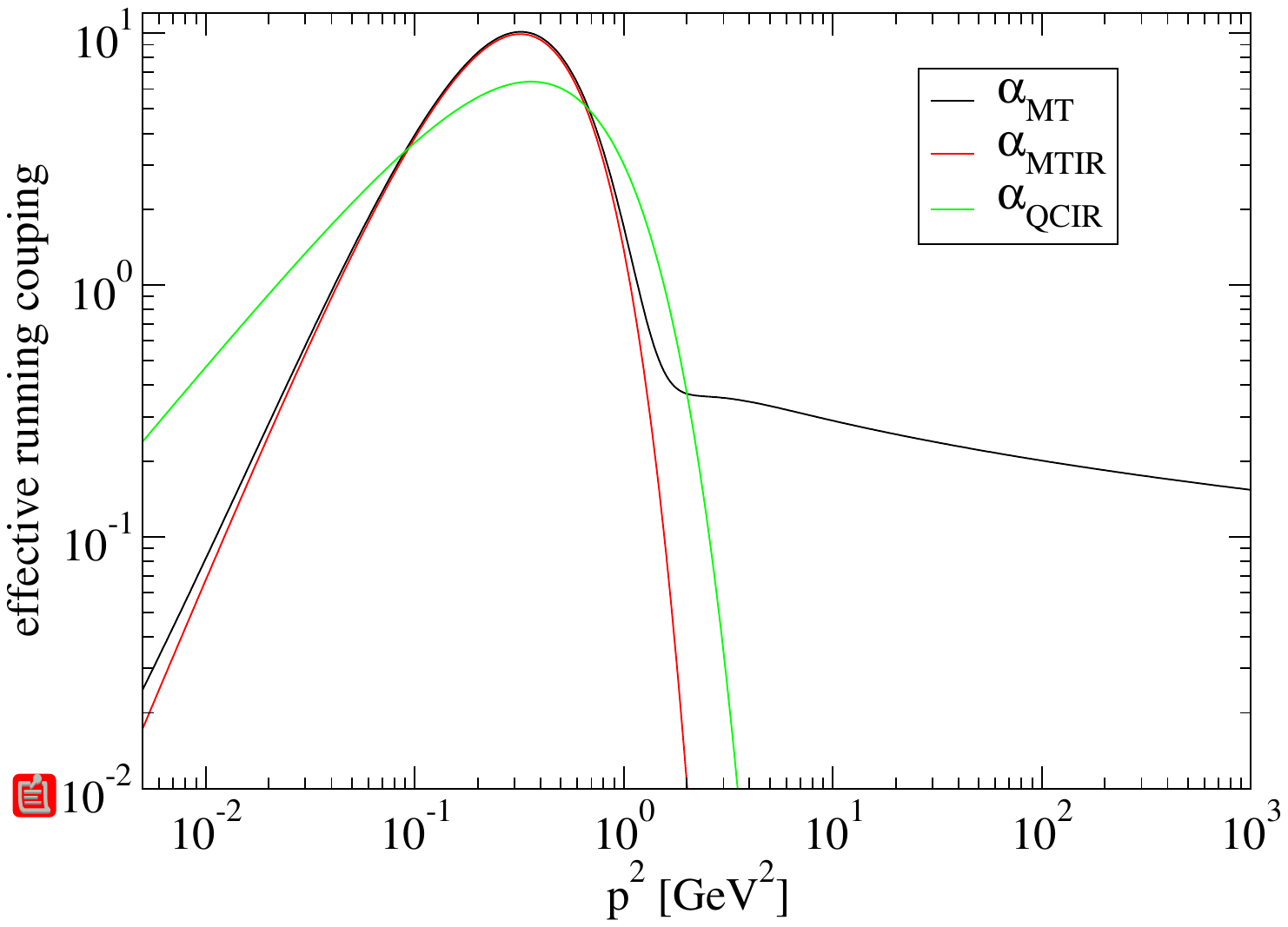}
	\caption{Comparison of our three models for the effective running coupling. See main text for details.}
	\label{fig:alpha}
    \end{figure}
    For later discussion we have plotted all three models in Fig.~\ref{fig:alpha}.
    In order to improve the models with \textit{some} unquenching physics and to account for the thermal splitting 
    of the gluon, we include a thermal mass for the gluon as in
    Ref.~\cite{Gao_2016}, i.e., for every $\alpha_\text{X}$ model, the longitudinal and transverse part of the effective running coupling 
    are taken according to
    \begin{equation}
	\begin{aligned}
		\alpha_L(q^2)&=\alpha_\text{X}\big((q^2+m_g(T,\mu)^2)\big),
		\quad \text{where} \quad	m_g(T,\mu)^2 &= \frac{16}{5}\left(T^2 +\frac{6\mu^2}{5\pi^2}\right),
		\\
		\alpha_T(q^2)&=\alpha_\text{X}(q^2).
	\end{aligned}
    \end{equation}
    The explicit form of the gluon mass $m_g$ is taken from HTL-HDL results, see Ref.~\cite{Thoma:1997bi,Haque:2012my}.
    The above is already far closer to QCD than the models described in the introduction. The interaction is a non-local and
    non-perturbative ``gluon'' exchange, as in QCD, and it reproduces important QCD properties such as multiplicative 
    renormalisability and in its full MT version asymptotic freedom.
	\begin{figure}
	\centering
	\begin{minipage}{0.32\linewidth}
		\begin{center}
			(a) Full Maris-Tandy
		\end{center}
		\includegraphics[width=\linewidth]{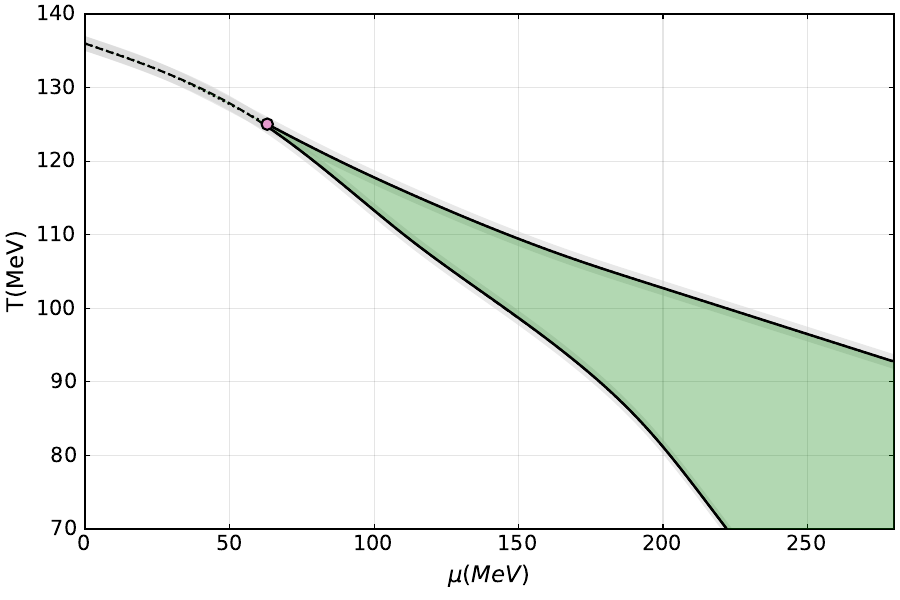}
	\end{minipage}
	\begin{minipage}{0.32\linewidth}
		\begin{center}
			(b) Maris-Tandy IR only
		\end{center}
		\includegraphics[width=\linewidth]{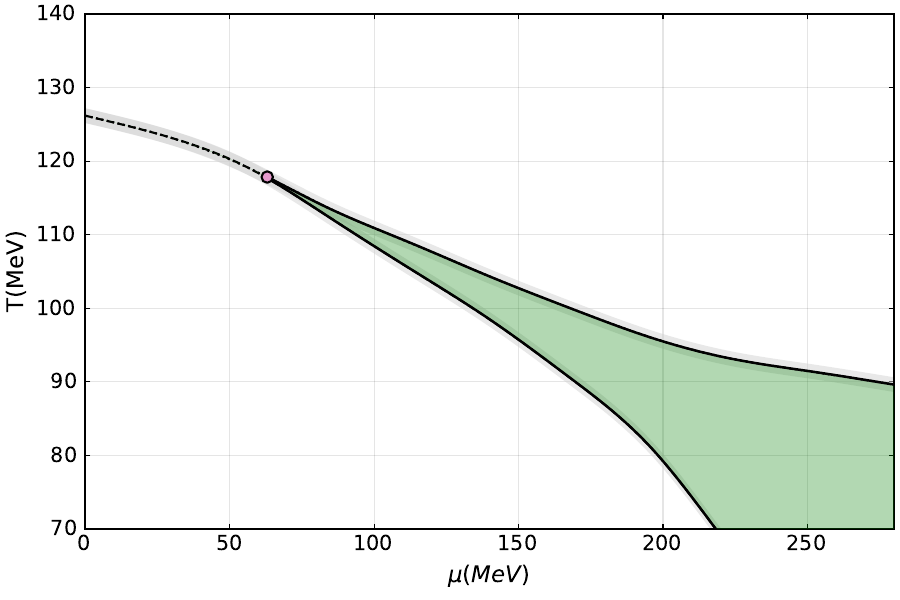}
	\end{minipage}
	\begin{minipage}{0.32\linewidth}
		\begin{center}
			(c) Qin-Chang IR only
		\end{center}
		\includegraphics[width=\linewidth]{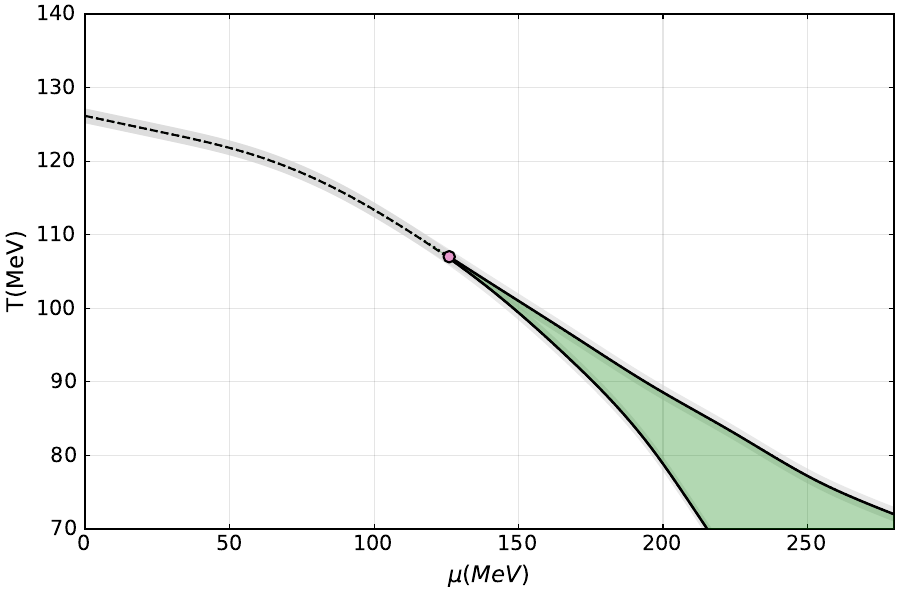}
	\end{minipage}
	\caption{Phase diagrams of the three models studied in this work, considering homogeneous phases only. The dashed line denotes a second-order phase transition, 
	the dot locates the tri-critical point, and the two lines which enclose the green region are the spinodals. The left 
	spinodal shows when the WW solution becomes stable (to the left of it, WW is unstable, to the right, it is stable or metastable), 
	and the right spinodal shows when the NG solution becomes unstable 
	(to the left the NG solution is stable or metastable, to the right it does no longer exist). In the green region both WW and NG are stable or metastable.}
    \label{fig:mypd}
    \end{figure}

    We may then calculate the homogeneous phase diagram of our models of QCD. 
    Searching for the stationary points of $\Gamma$, i.e.,
    \begin{equation}
	\frac{\delta \Gamma}{\delta S} = 0,
    \end{equation}
    we obtain the quark propagator (qDSE), which is diagrammatically given by
    \begin{equation}\label{qDSE}
	\centering%
	\begin{tikzpicture}
		\begin{feynman}
			\vertex (a);
			\vertex [right=of a] (c);
			\vertex [right=0.5cm of c] (d);
			\vertex [above=0.4cm of c] (m1);
			\diagram*{
				(a) -- [fermion3] (c)
			};
			\draw (d) node { \({=}\)};
			\draw (m1) node { \(^{-1}\)};
			\vertex [right=0.5cm of d] (a1);
			\vertex [right=of a1] (c1);
			\vertex [right=0.5cm of c1] (d1);
			\vertex [above=0.4cm of c1] (m12);
			\diagram*{
				(a1) -- [fermion2] (c1)
			};
			\draw (d1) node { \({+}\)};
			\draw (m12) node { \(^{-1}\)};
			\vertex [right=0.5cm of d1] (a2);
			\vertex [right=0.55cm of a2] (gl1);
			\vertex [right=1cm of a2] (b2);
			\vertex [above=0.4cm of b2] (gldot);
			\vertex [right=1cm of b2] (c2);
			\vertex [left=0.5cm of c2] (gl2);
			\diagram*{
				(a2) -- [fermion2] (b2) -- [fermion2] (c2);
				(gl1) -- [boson,half left] (gl2)
			};
			\draw (gldot) node [gray, dot];
			\draw (b2) node [dot];
			\draw (gl2) node [];
		\end{feynman}
	\end{tikzpicture}\,.
    \end{equation}
    For homogeneous solutions, it explicitly reads
    \begin{equation}
	\left[\bar{S}(k)\right]^{-1} = Z_{2}\left[S_0(k)\right]^{-1} 
	+ C_{F}\,Z_{2}^2 \,\SumInt\, 
	\gamma_\mu \,\bar{S}(q)\, \gamma_\nu\, \left(P^T_{\mu\nu}(l) \frac{\alpha_T(l^2)}{l^2}+ P^L_{\mu\nu}(l) \frac{\alpha_L(l^2)}{l^2}  \right)_{l=q-k}\,,\label{DSEs-1} 
    \end{equation}
    where $C_F=4/3$ is the Casimir factor.
    We solve this equation for multiple values of temperature $T$ and chemical potential $\mu$. The value of the chiral condensate 
    $\Delta \propto \Tr{[S]}$ can be used as an order parameter to compute a full phase diagram. We display our results for all three
    models in Fig.~\ref{fig:mypd}.
    In this paper, we consider two quark flavours and,
    as mentioned earlier, we work in the chiral limit for simplicity. 
    We then encounter a second-order chiral transition for small chemical potential and large temperatures,
    which turn into a tri-critical point and a first order transition when the chemical potential is increased. The spinodal 
    region inside which the first order transition occurs is shaded in green. We observe that all three models display a 
    critical point at rather small (quark) chemical potential. This is a generic feature of models of this type as discussed, e.g.,
    in Ref.~\cite{Fischer:2018sdj}.

	\section{Results}\label{sec:results}
	
	\subsection{Cross-check: instability towards homogeneous broken phase}\label{sec:crosscheck}

	\begin{figure}
	\centering
	\begin{minipage}{0.32\linewidth}
		\begin{center}
			(a) Full Maris-Tandy
		\end{center}
		\includegraphics[width=\linewidth]{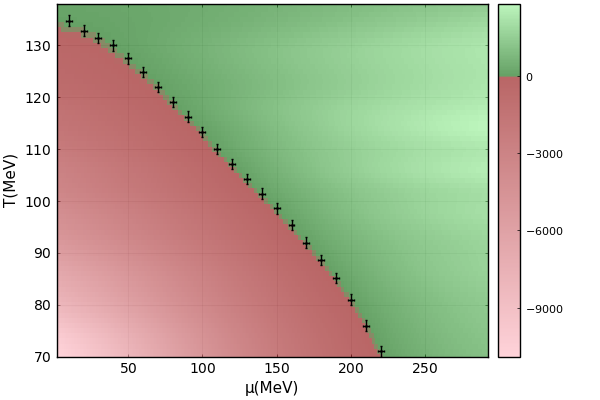}
		\includegraphics[width=\linewidth]{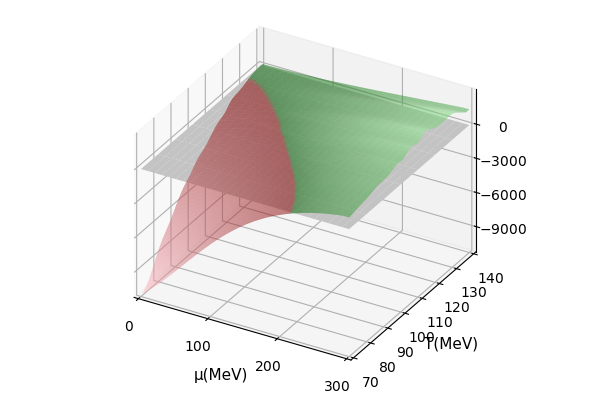}
	\end{minipage}
	\begin{minipage}{0.32\linewidth}
		\begin{center}
			(b) Maris-Tandy IR only
		\end{center}
		\includegraphics[width=\linewidth]{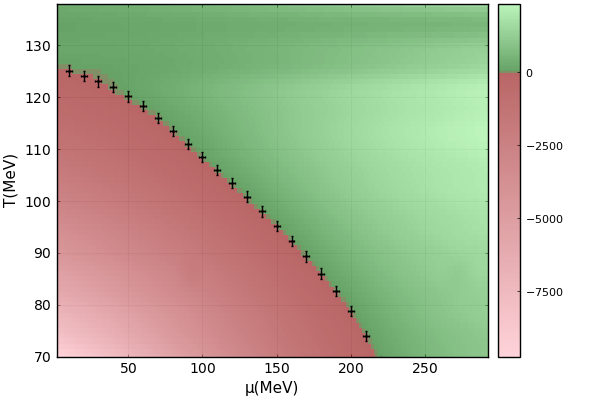}
		\includegraphics[width=\linewidth]{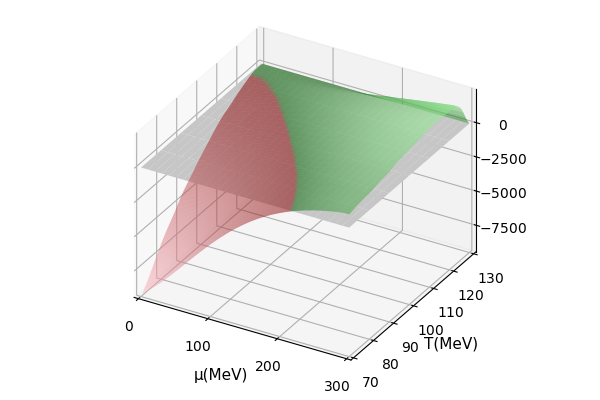}
	\end{minipage}
	\begin{minipage}{0.32\linewidth}
		\begin{center}
			(c) Qin-Chang IR only
		\end{center}
		\includegraphics[width=\linewidth]{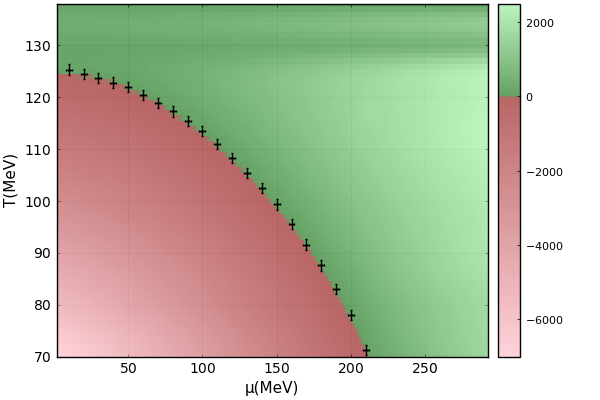}
		\includegraphics[width=\linewidth]{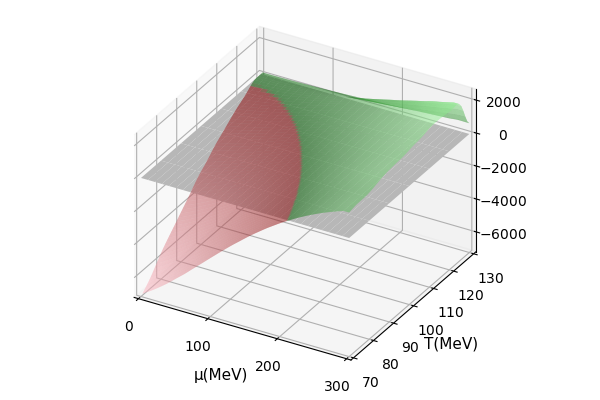}
	\end{minipage}
	
	\caption{Stability analysis of the homogeneous chiral phase. Red regions correspond to negative $\tilde \Omega^{(2)}(0)$ at the saddle point, meaning that the WW solution is unstable there. The crosses indicate the 
    (meta-) stability boundary of the  WW solution from the explicit calculations of the qDSE in Fig.~\ref{fig:alpha},
    i.e., the second-order phase boundary or the left spinodal. Both the colour axis on the heatmaps and z-axis in the surface plots correspond to $\tilde \Omega^{(2)}(0)$ in arbitrary units.
    }
	\label{fig:ct}
    \end{figure}
    Having specified our approximation scheme, we are now in a position to discuss the results of our stability
    analyses.
    We first report on specific cross-checks, analysing the instability of the chirally symmetric solution 
    with respect to \textit{homogeneous} chiral-symmetry breaking perturbations. 
    More precisely, we test whether with our setup we can reproduce the boundaries where in Fig.~\ref{fig:mypd} the homogeneous WW solution becomes unstable, i.e., the second-order phase boundary at high temperatures and the
    left spinodal at lower temperatures. Passing this test is a necessary condition to ensure the number of parameters in the test-function is sufficiently big. We already did this in Ref.~\cite{Motta:2023pks}. 
    However, since we now consider different model interactions and, in particular, since we extend our studies to
    lower temperatures, it is important to repeat the test.   
    
    To this end we evaluate Eq.~(\ref{eq:Omt2}) 
    for $d=0$, which then simplifies to
     \begin{equation}\label{eq:Omt0}
    \tilde \Omega^{(2)}(0)=
    -\SumInt_{k} \text{tr}\Big[
    \bar{S}(k)\epsilon_m(k)
    \bar{S}(k)\epsilon_m(k)
    \Big]
    -g^2Z_{1F}\SumInt_{kl}
    \text{tr}\Big[\gamma_\mu t^a  \bar{S}(k)
    \epsilon_m(k)
    \bar{S}(k)
    \gamma_\nu t^b
    \bar{S}(l)
    \epsilon_m(l)
    \bar{S}(l) D_{\mu\nu}^{ab}(k-l)	\Gamma_{qg}(k-l)	\Big],
    \end{equation}    
    minimise it with respect to the parameters 
    $L_1$ and $L_2$, maximise with respect to $L_{3,4,5}$,\footnote{    
    Typical values of $L_{1,2,4,5}$ at the saddle point range from 600 to 800 MeV and the dimensionless $L_3$ is typically around -1/2.} 
    and then check its sign.
    Our results are shown in
	Fig.~\ref{fig:ct}. Indeed we find that our generalised test function, Eq.~(\ref{epsm}), is perfectly adequate. Had we restricted
	ourselves, however, to the original function defined in Ref.~\cite{Motta:2023pks} we would have failed in the low temperature 
	region. More precisely, with an inadequate test-function one can find spurious instabilities.
 Thus, we have arrived at a big enough parameter space to adapt to the complex mathematical structure of the stability criterion\footnote{Note that the integrands of Eq.~(\ref{eq:Omt0}) depend on the homogeneous solutions for the qDSE. Together with our definition of $\epsilon_m$ this becomes rather complicated. 
		As we lower the temperature, 
	the lowest Matsubara frequency gets smaller and the IR part of the integrand becomes more relevant. This complexity is what, according to our understanding, forces us to increase the number of parameters in our test-function.} and correctly find the saddle point.

	\subsection{Instability towards inhomogeneous broken phase}\label{sec:inhom}

	With the cross-check successfully passed, we then use the same form of $\epsilon_m$ in the test function $\epsilon_\Sigma(k_1,k_2)$ for the inhomogeneous case, Eq.~(\ref{test-functionSig}). 
	Unfortunately, away from the homogeneous $k_1=k_2$ limit the computations get severely more time-intensive. Furthermore, 
	for large momentum asymmetry, one might wonder whether or not our form for $\epsilon_m$ in Eq.~(\ref{epsm}) still has 
	enough parameters, and without an analogue of the test in Fig.~\ref{fig:ct}, one would need to run extensive tests to 
	ensure the sufficiency of the parameter space of the test-function. 
	Therefore, we restrict our analysis in this work to points on the left spinodal line, for which we know from the test in Fig.~\ref{fig:ct} that our form of $\epsilon_m$ is adequate. Going along this line we perform the stability analysis for several values of $d\equiv|\vec k_1-\vec k_2|$. (As said above, we only consider static perturbations, i.e., vanishing energy differences $d_4=\omega_1-\omega_2=0$.)

\begin{figure}
		\centering
		\includegraphics[width=1\linewidth]{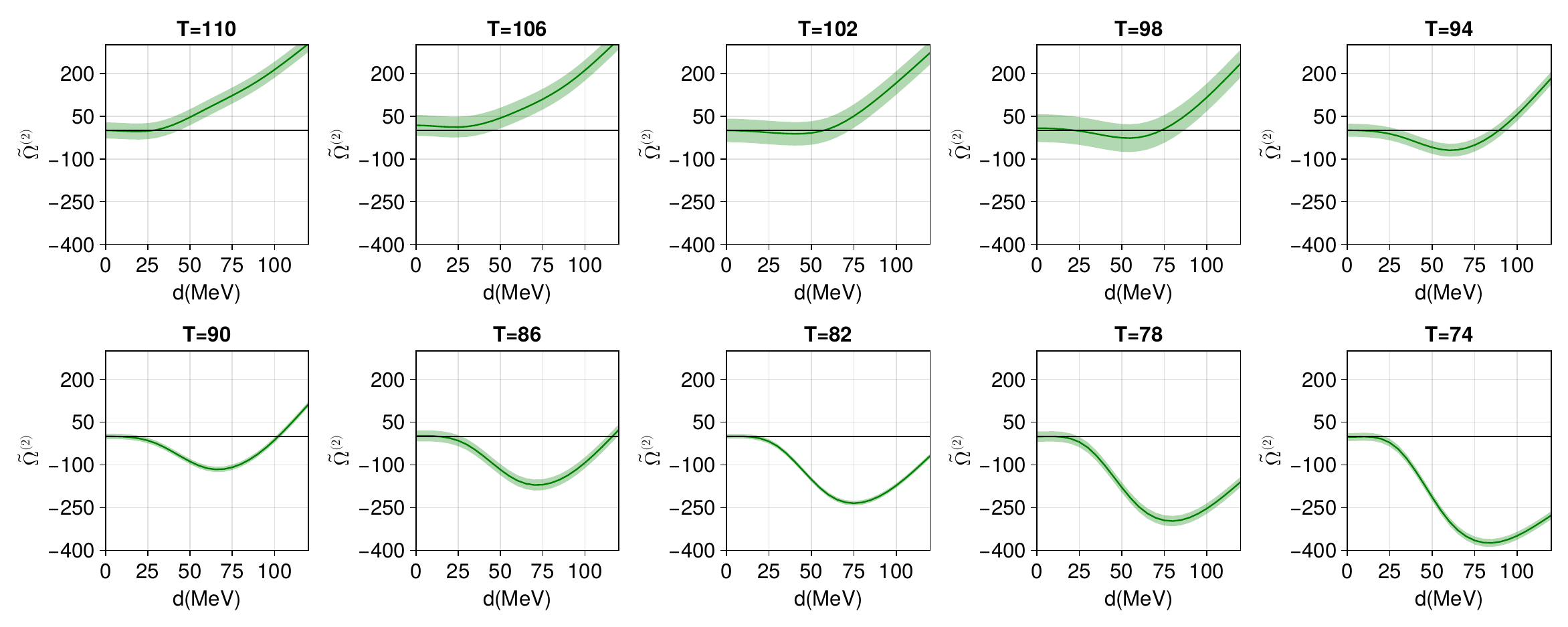}
		\caption{Stability analysis of the WW phase for points of different temperatures on the left spinodal of the MT model (see the homogeneous phase diagram in Fig.~\ref{fig:mypd}).
        $\tilde \Omega^{(2)}$ is plotted as a function of the difference $d = |\vec k_1 - \vec k_2|$ of the two momenta in the test function Eq.~(\ref{test-functionSig}).}
		\label{fig:fullsa}
	\end{figure}

    Our results for the MT model are shown in Fig.~\ref{fig:fullsa} for different temperatures. Since we are on the spinodal, $\tilde \Omega^{(2)}$ vanishes in the homogeneous limit $d=0$ within numerical errors.\footnote{The error bands on $\tilde \Omega^{(2)}(d)$ are a conservative estimate of the error due to the fact that the calculation is performed on a grid in $\mu$ and thus we never sit exactly \textit{on} the spinodal line but rather slightly above.}
    Going away from this limit, we see that for large temperatures, $\tilde \Omega^{(2)}$ increases with increasing $d$, meaning that the homogeneous WW solution is stable against small inhomogeneous perturbations.
    However, below a certain temperature, this is no longer the case. Here we see that $\tilde \Omega^{(2)}$ is negative in a certain regime of $d>0$, indicating that the WW solution now is \textit{unstable} against small inhomogeneous perturbations. More precisesly, $\tilde \Omega^{(2)}$ first decreases with increasing $d$, reaches a minimum and increases again at higher $d$.     
	This moat-shaped curve is expected, as we should not find instabilities against
    arbitrarily large momentum asymmetries. 
	
    However, we should keep in mind that we performed our stability analysis on the WW solution, which on the left spinodal is disfavored against the homogeneous NG solution. Therefore the fact that the WW phase is unstable against inhomogeneous perturbations does not necessarily imply that there is indeed an inhomogeneous ground state at those points. 
    This conclusion would only be valid in the region where the WW solution is favored over the NG solution, i.e., at chemical potentials beyond the homogeneous first-order phase boundary.     
    We have not determined the exact location of this line from the pressure difference but it must lie somewhere between the two spinodal lines. 
    We note that the effect of increasing the chemical potential is basically to 
	lift the value of $\tilde\Omega^{(2)}$ without changing much the functional form of the plots in Fig.~\ref{fig:fullsa}. Hence, since we find that on the left spinodal the well in $\tilde\Omega^{(2)}(d)$ gets deeper with decreasing temperature, we also expect the instability to persist for a longer chemical-potential interval, eventually perhaps beyond the homogeneous first-order phase boundary. We will come back to this later on. 
 
	As already seen in Fig.~\ref{fig:mypd}, the MT model features a Tri-Critical Point (TCP) roughly at 
	\begin{equation}
	(T^{TCP},\mu_q^{TPC})=(125,58) \mbox{ MeV}\,.
	\end{equation} 
    As we can see from the results in Fig.~\ref{fig:fullsa}, the temperature for which the WW phase starts to become unstable 
    with respect to crystallisation is at roughly 100 MeV.
    This is substantially lower in temperature than the TCP. If this point was located on the second-order chiral phase boundary \textit{above} the TCP, it would presumably be a Lifshitz point (LP), i.e., a point in which three second-order lines meet, separating a homogeneous broken phase, a homogeneous symmetric phase, and an inhomogeneously broken phase. However, since the point lies below the TCP on the left spinodal, i.e., in a region where homogeneous NG solution is favored over the WW solution we will refer to it as a proto-LP point.
	
	\begin{figure}[t]
		\centering
		\includegraphics[width=0.495\linewidth]{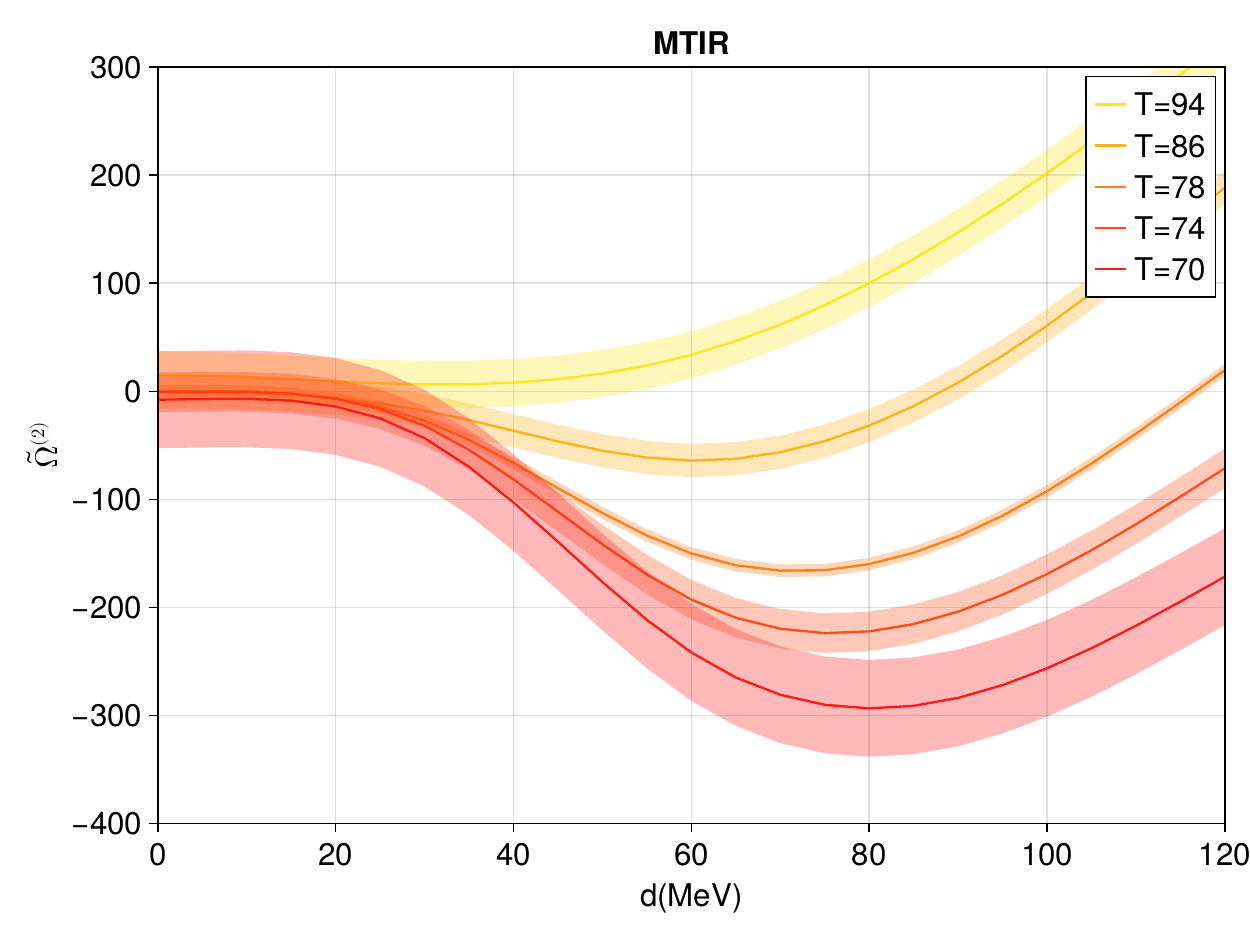}
		\includegraphics[width=0.495\linewidth]{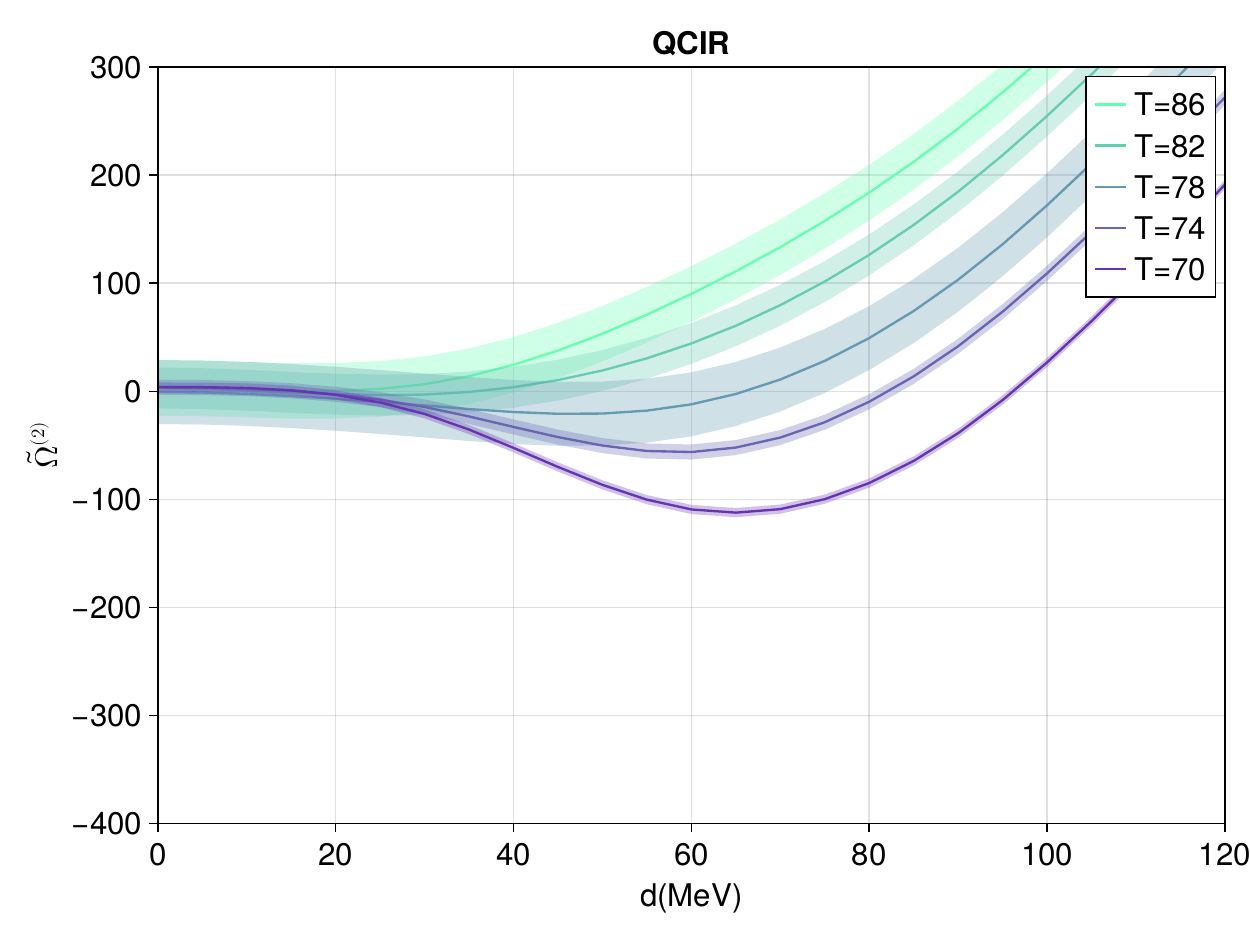}
		\caption{The same as Fig.~\ref{fig:fullsa} but for the MTIR model (left) and the QCIR model (right).}
		\label{fig:slices}
	\end{figure}

    Before we discuss the potential physical meaning of this point, it is instructive to compare our result for the full MT-model with results for the two other models for the effective running coupling, MTIR and QCIR.  
    The results for these models are displayed in Fig.~\ref{fig:slices}. In the left panel we find that at a temperature of $94$ MeV, where the full MT model already shows an instability, the MTIR is still stable. For the QCIR model (right panel), the onset of instability is moved even further down to lower temperatures. 
    However, if we compare this with the homogeneous phase diagrams shown in Fig.~\ref{fig:mypd}, there seems to be some correlation between the location of the TCP and that of the proto-LP. The higher the TCP, the higher the proto-LP, for all models investigated.

	\section{Discussion}\label{sec:discussion}

\begin{figure}[t]
	\centering
	\includegraphics[width=0.48\linewidth]{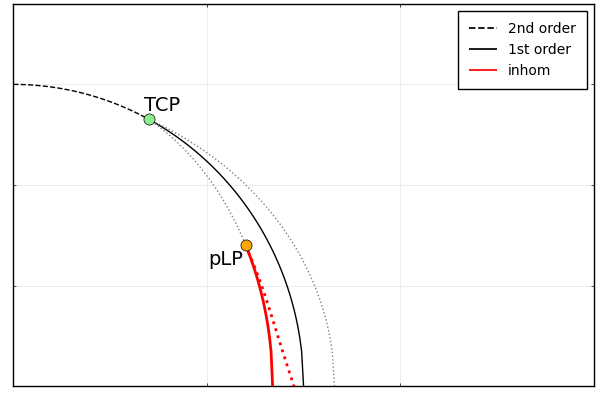}
	\raisebox{0cm}{\includegraphics[width=0.48\linewidth]{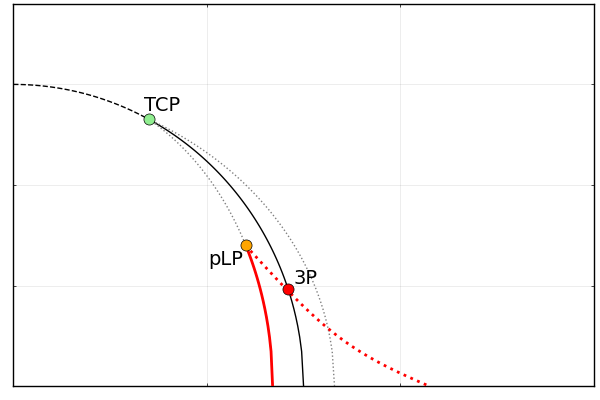}}
	\caption{Sketches of the two possible outcomes in all models discussed in this manuscript and possibly in full QCD. The solid red line represents the region where we do find an instability of the WW solution against inhomogeneous perturbations and the dotted red line represents the high-$\mu$ boundary of the instability region. In the left panel this line does not cross the homogeneous first-order phase boundary (black solid line), while in the right panel it does at a triple point (3P). In the latter case there must be an inhomogeneous phase below the 3P between the black solid and red dotted lines.}
	\label{fig:sketch}
\end{figure}

    As pointed out above, the instabilities of the WW solutions on the left spinodal against inhomogeneous perturbations do not necessarily imply the existence of an inhomogeneous phase since we do not know how far in $\mu$ these instabilities go for a given temperature.
    However, we can say with certainty that they are there.
    Moreover, as the dip gets deeper it is likely that the lower the temperature, the farther they will stretch out in $\mu$. Therefore, there are two possible outcomes.
	Either this instability of the homogeneous WW phase persists at high enough values of $\mu$, past the first-order phase 
	transition line, where the WW solution is the energetically favoured homogeneous solution. Or it does not. If it does not, then this instability is artificial, and it would never be realised in nature. However, if it does persist past the first order transition line, 
	then the most stable homogeneous phase is unstable against inhomogeneities, i.e., the ground state must be inhomogeneous. 
 These two
	possibilities are shown in Fig.~\ref{fig:sketch}. If the red dotted line in Fig.~\ref{fig:sketch}, which represents the 
	onset of the instability of the WW phase from higher chemical potentials, goes over the homogeneous first-order transition line, depicted as the solid black curve, a
	Triple-Point is formed, where the homogeneous WW and NG phases and an inhomogeneous phase meet.
 
	The two possibilities shown in Fig.~\ref{fig:sketch} are the \textit{most conservative} interpretation of our results. 
	Note, that it could very well be that for temperatures higher than the proto-LP shown in the sketch, the WW solution 
	is not unstable with respect to the introduction of a \textit{small} inhomogeneous perturbation, but metastable, and 
	large disturbances could trigger its transition to a crystalline phase. This is something we are not ruling out. Our 
	analysis is only sensitive to unstable solutions. It should also be said that a truly exhaustive analysis would need 
	to be performed with several different test functions. Both the functional form for $\epsilon_m$ should be varied, but 
	also $\epsilon_\Sigma$ could in principle be changed. Our analysis is \textit{inclusive}, rather than \textit{exclusive}, 
	meaning that whenever it sees an instability, we can be sure one exists, however, if it does \textit{not} see an instability, 
	no claim can be made. We chose to not perform this most expensive study with different test functions, and going beyond 
	the WW phase boundary, due to the fact that these are simple models. It makes little sense putting massive effort to 
	understand the Maris-Tandy model's phase diagram. This 
	more in-depth analysis only make sense, to our mind, for truncations even closer to QCD.
  
	It is noteworthy, however, that our result is very similar to what is seen in non-local-NJL calculations.
	Refs.~\cite{Carlomagno:2014hoa,Carlomagno:2015nsa}, for instance, perform their calculations in such a setup. The model's 
	non-locality is introduced as a form factor on the quark-quark contact interaction, which is different from what 
	we do here. However, in some sense, the models we discuss here are similar to non-local NJL models, since the gluons 
	are not dynamic. It is interesting to note this consistency and the phase diagrams we can extrapolate from our results (Fig.~\ref{fig:sketch}) are similar to those in Ref.~\cite{Carlomagno:2015nsa}.

    Finally, there is one subtle point, made in Ref.~\cite{Koenigstein:2021llr} that we find is important to point out. In the GN model, where the full analytic solution is known, 
    there is an inhomogeneous region slightly to the left of the homogeneous first-order transition, where the 
    homogeneous broken solution is preferred over the symmetric solution.
    In this region, Ref.~\cite{Koenigstein:2021llr} shows that if one performs a stability analysis of the broken solution, it is found to be stable, and the symmetric solution is unstable. However, the full analytic solution shows the ground state to be inhomogeneous. They discuss the emergence of a ``potential barrier'' that causes the homogeneous broken phase to become stable against inhomogeneous perturbation in this region, even though the broken phase is preferred over the symmetric one and even though the true analytic solution shows an inhomogeneous phase.

	\section{Conclusions and Outlook}\label{sec:conclusions}
     
	We have applied the framework introduced in Ref.~\cite{Motta:2023pks} for the first time to search for crystalline phases. Our findings within Rainbow-Ladder truncation with modelled running coupling are positive, an instability towards inhomogeneous breaking of chiral symmetry does exist, the WW solution is unstable against translational symmetry breaking. However, in order to really guarantee that the phase is realised at larger chemical potentials, two extra ingredients are necessary. We need a clear determination of the first-order phase transition line, and, since we do not have an analogue of the homogeneous chiral phase boundary that we can use to further guarantee the parameter space of our test-functions is sufficiently big, a more systematic search is warranted.
	
	Nevertheless, we point out that the tendency towards crystallisation that is seen in NJL and QM models, also exists for the models we studied here. We intend to give a clearer determination of whether or not these phases play a role in high density physics on a future publication, by use of a truncation that takes us much closer to QCD than simple models.

	Furthermore, it is important we state that the truncation and models we studied in this paper do not capture all relevant physics.
    It is well established in the literature that quantum fluctuations might disorder inhomogeneous phases \cite{Winstel:2024qle,Pisarski:2020dnx,Hidaka:2015xza,Lee:2015bva}, in particular fluctuations by the Nambu-Goldstone bosons associated with the breaking of translational symmetry (the phonons, leading to the so-called Landau-Peierls instability) and flavour symmetry (the pions, see \cite{Pisarski:2020dnx}). If these phases are disordered by fluctuations, they do not, however, go back to a normal homogeneous phase. Rather, they become either liquid-crystals or the so-called Quantum Pion Liquid (QPL). What a stability analysis can then say is that there most certainly is something going on in this region. It could be an inhomogeneous phase, or it could be some other exotic phase such as liquid-crystals or a QPL. It is also known that explicit breaking of isotropy might contribute to the stabilisation of the crystalline phase against Landau-Peierls instability. The presence of magnetic fields for instance, in collusion with topological effects (see Refs.~\cite{Ferrer:2021mpq,Ferrer:2019zfp}) makes these phases stable with respect to thermal fluctuations. 
    The machinery that we use in this paper, namely, the 2PI stability analysis, can be applied to any truncation, including those which take into account Nambu-Goldstone boson fluctuations dynamically.
	
	\section*{Acknowledgements}
	It is our pleasure to thank Fabian Rennecke, Laurin Pannullo, Marc Winstel, Wilhelm Kroshinsky, Lorenz von Smekal, and Peter Lowdon for helpful discussions about the content of this manuscript. 
    This work has been supported by the Alexander von Humboldt Foundation, the Deutsche Forschungsgemeinschaft (DFG) through 
    the Collaborative Research Center TransRegio CRC-TR 211 ``Strong-interaction matter under extreme conditions'' and the 
    individual grant FI 970/16-1, the Helmholtz Graduate School for Hadron and Ion Research (HGS-HIRe) for FAIR and the GSI
    Helmholtzzentrum f\"{u}r Schwerionenforschung. 
	
	\bibliographystyle{ieeetr}
	\bibliography{stabilitybib.bib}
	
\end{document}